# Probing photo-induced rearrangements in the NdNiO$_3$ magnetic spiral with polarization-sensitive ultrafast resonant soft X-ray scattering

K.R. Beyerlein[1,2], A.S. Disa[1,2,3], M Först[1,2], M. Henstridge[1,2], T. Gebert[1,2], T. Forrest[4], A. Fitzpatrick[4], C. Dominguez[5], J. Fowlie[5], M. Gibert[6], J.-M. Triscone[5], S.S. Dhesi[4] and A. Cavalleri[1,2,3,7]

[1]Max Plank Institute for the Structure and Dynamics of Matter, Hamburg, Germany
[2]Center for Free-Electron Laser Science, Hamburg, Germany
[3]The Hamburg Centre for Ultrafast Imaging, Hamburg, Germany
[4]Diamond Light Source, Harwell Science and Innovation Campus, Didcot, UK
[5]Department of Quantum Matter Physics, University of Geneva, Geneva, Switzerland
[6]Department of Physics, University of Zurich, Zurich, Switzerland
[7]Department of Physics, Clarendon Laboratory, University of Oxford, UK

ABSTRACT

We use resonant soft X-ray diffraction to track the photo-induced dynamics of the antiferromagnetic structure in a NdNiO$_3$ thin film. Femtosecond laser pulses with a photon energy of 0.61 eV, resonant with electron transfer between long-bond and short-bond nickel sites, are used to excite the material and drive an ultrafast insulator-metal transition. Polarization sensitive soft X-ray diffraction, resonant to the nickel $L_3$-edge, then probes the evolution of the underlying magnetic spiral as a function of time delay with 80 picosecond time resolution. By modelling the azimuthal dependence of the scattered intensity for different linear X-ray polarizations, we benchmark the changes of the local magnetic moments and the spin alignment. The measured changes are consistent with a reduction of the long-bond site magnetic moments and an alignment of the spins towards a more collinear structure at early time delays.

Rare-earth nickelates exhibit sharp metal-insulator transitions and unusual noncollinear antiferromagnetic (AFM) spin-ordering at low temperatures [1], which may become interesting for device applications [2]. Amongst these nickelate compounds, we analyze $NdNiO_3$ – a material that exhibits exceptionally strong coupling of structural, electronic, and spin degrees of freedom. Despite previous indications of charge disproportionation [3–8], recent measurements suggest that the ground state of $NdNiO_3$ is a bond-disproportionate, negative charge transfer insulator [9,10]. In this model, the nickel atoms have a $Ni^{2+}$ valence state ($d^8$) at all temperatures, strongly hybridizing with the oxygens in the surrounding octahedra by taking an electron and leaving behind a hole on the ligand ($L$) [11,12]. In the high temperature metallic state, the oxygen holes are equally distributed among all nickel sites resulting in a $d^8\underline{L}$ electronic configuration. The low temperature structure contains two different sized oxygen octahedra, with either long (LB) or short Ni-O bonds (SB), that are ordered in a checkerboard pattern. In the limit of this bond-disproportionate description, the electronic configurations for these two nickel sites are $d^8$ (LB) and $d^8\underline{L}^2$ (SB).

The nickel ions ($t_{2g}^6 e_g^2$) retain a high spin state (S=1), which is partially screened by the oxygen holes. In the metallic phase, this screening leads to a total spin of around 0.9 $\mu_B$ on each Ni octahedron. In the insulating phase, screening depends on the Ni-O bond length with higher spin found on the LB-sites (1.3 $\mu_B$) than on the SB-sites (0.7 $\mu_B$) [12]. Furthermore, the spins order into an antiferromagnetic structure with a supercell consisting of a sequence of four (111) lattice planes with different spin orientations. Powder neutron diffraction measurements indicated a collinear AFM structure of the spins (↑↑↓↓) [13,14], while polarization sensitive resonant soft X-ray scattering [15–17] and inelastic X-ray scattering [18] on thin films have found a non-collinear spin spiral structure (↑→↓←) . Recent X-ray scattering studies on thin films have also found the spins to have a canting angle less than 90°, with a collinear structure emerging in the limit of monolayer thickness [19,20].

Optical pulses have been shown to dynamically induce transitions between insulating and metallic phases in $NdNiO_3$ thin films, using either light resonant with optical phonon vibrations in the underlying substrate [21–24] or by exciting electrons into the conduction band of the nickelate film directly [25–28]. In some of these studies, time-resolved resonant soft X-ray diffraction was used to probe the evolution of the magnetic order parameter across these transformations by measuring transient changes in the intensity of a superlattice reflection for a fixed incident X-ray polarization. Further details of the magnetic structural dynamics can be obtained by extending modern X-ray scattering methods to the time domain. For instance, analyzing the polarization dependence of the scattered intensity at each time delay can provide information about the relative spin moments and orientations on different sites [15–17]. Here, we combine time and polarization resolved resonant soft X-ray diffraction to provide new insight into the dynamical changes of the AFM structure across the light-induced phase transformation, as well as, during its recovery back to the equilibrium state.

A thin film of NdNiO$_3$ with a pseudo-cubic [111]$_{pc}$ surface normal was grown on a LaAlO$_3$ substrate following previously detailed methods [29,30]. This produced a uniform film of 36 nm thickness that was coherently strained to the substrate, as discussed in Appendix A.

The NdNiO$_3$ thin film was excited by femtosecond light pulses with a photon energy of 0.61 eV (2 $\mu$m wavelength) and its magnetic state was probed using time-resolved resonant soft X-ray scattering at the I06 beamline of the Diamond Light Source. An illustration of the experimental scattering geometry is shown in Figure 1. This excitation photon energy was chosen to study dynamics associated with transitions close to the band gap, as previous studies have utilized a higher energy of 1.55 eV (800 nm wavelength) [25–28]. Ellipsometry measurements suggest that such a near band gap excitation promotes inter-site electron transitions from occupied states with LB-site character to unoccupied hybridized Ni 3d – O 2p states primarily on SB-sites [31–34].

The ¼ ¼ ¼ magnetic Bragg peak intensity dependence on incident X-ray polarization and sample azimuthal angle was measured to determine the magnetic structure as a function of time delay. The X-ray pulses were produced using the hybrid filling mode of the storage ring, consisting of a contiguous train of electron bunches followed by a single bunch with a higher charge of 3 nC, which was separated by 250 ns from the bunch train. A pair of APPLE-II undulators were then used to generate soft X-ray pulses of ~80 picosecond duration with control over the linear polarization direction [35]. A grating monochromator then selected an X-ray photon energy of 852 eV with 0.1 eV bandwidth, which is resonant with the nickel $L_3$-edge. The sample was cooled to 40 K on a cryogenic sample stage, stabilizing an equilibrium AFM insulating state in the NdNiO$_3$ film. The scattered intensity was measured in a reflection geometry using a micro-channel plate (MCP) detector. The magnetic Bragg peak intensity dependence on the incident X-ray energy near the nickel $L_3$-edge was also measured and found to be consistent with previous reports [25].

For optical excitation, the output of a 5 kHz Ti:sapphire amplifier system pumped an optical parametric amplifier to generate 2 $\mu$m near-infrared light pulses of 80 $\mu$J pulse energy. These pulses were focused onto the sample with a spot size of 350 $\mu$m corresponding to an excitation fluence of about 16 mJ/cm$^2$. As shown in Figure 1, the near-IR beam path was offset 30 degrees in the scattering plane with respect to the incident X-ray beam. The excitation pulses were synchronized to the ~535 kHz train of isolated high bunch charge X-ray pulses, and laser-on and laser-off events were selectively detected by gating the readout signal of the MCP detector. The pump-probe time delay was controlled electronically via the synchronization system.

For each azimuthal orientation angle ($\Psi$) of the sample, the peak scattered intensity as a function of pump-probe time delay was measured first using X-rays with linear polarization aligned parallel to the scattering plane ($\pi$-polarization), and then using a linear polarization perpendicular to the plane ($\sigma$-polarization). These time-resolved measurements were completed for the sample orientations $\Psi = -45°$, $0°$, $45°$, $90°$ and $120°$. Laser-off measurements were made for this set of orientations as well as for $\Psi = -90°$. We have defined the angle $\Psi$ to be consistent with Ref. 19. Post-processing of the data consisted of correcting for dark measurements taken before each scan and renormalizing the integrated intensity to the incident X-ray flux.

First, the equilibrium magnetic structure of the $NdNiO_3$ film was determined from the measured laser-off data. The ratio of the scattered intensities from $\pi$- and $\sigma$-polarized incident X-rays ($I_\pi / I_\sigma$) as a function of $\Psi$ is shown in Figure 2a. The magnetic structure was refined from these data assuming a unit cell consisting of four (111) scattering planes with variable moment magnitude and canting angle. Details of this modelling are described in Appendix B.

The spin-disproportionate, canted AFM magnetic structure, shown in Figure 2b, was found to best match the equilibrium data. The refined moments of 1.3 $\mu$B and 0.7 $\mu$B for the LB and SB sites, as well as the LB-site canting angle of 75 degrees relative to the (111) direction, agree well with earlier measurements of $NdNiO_3$ films of a similar thickness [19]. These values were determined with uncertainties of 0.2 $\mu$B and 3 degrees, respectively, suggesting that the number of azimuthal angles measured were enough to reliably refine the equilibrium structure.

Next, we turn to the evolution of the magnetic structure after optical excitation. The left panel of Figure 3 shows $I_\pi$ and $I_\sigma$ of the ¼ ¼ ¼ magnetic Bragg reflection, measured as a function of pump probe time delay for the set of measured $\Psi$ orientations. The melting of the AFM order in the $NdNiO_3$ film is seen from the sharp decrease in the scattered intensities for both polarizations near time zero. These intensity drops are followed by a gradual recovery on a time scale of 10 nanoseconds. However, the magnitude of the intensity drop and the recovery time depend strongly on $\Psi$ and the incident X-ray polarization, suggesting a change of the local magnetic structure with time after the excitation. We find the intensity drop to be larger for $\pi$-polarized than for $\sigma$-polarized X-rays, probably due to the higher sensitivity of this polarization to the noncollinear magnetic order of $NdNiO_3$ [19]. In particular, for $\Psi = \pm 45°$, $I_\sigma$ did not show a significant change after the optical excitation. At this azimuthal angle, noncollinear and collinear AFM structures of equilibrium NdNiO$_3$ films with different thicknesses have also shown negligible difference in $I_\sigma$ [19].

The time-resolved intensity changes for all sample azimuthal angles and X-ray polarizations were then fit to the product of an exponential and an error-function using the non-linear least squares refinement package LMFIT [36]. The resulting fits and 95% confidence intervals are shown together with the data in the left panel of Figure 3. The differences in temporal response for each azimuthal angle become more evident from the changes in $I_\pi / I_\sigma$ shown in the right panel of Figure 3. A sharp decrease in this ratio shortly after time zero is found in most cases, however, the minimum value and recovery time is different for each angle. The change in $I_\pi / I_\sigma$ observed for $\Psi = 120°$ is very small, indicating that this azimuthal angle is only weakly sensitive to the light-induced magnetic structure perturbation. This is also seen in the simulations that we will discuss later to extract the changes in magnetic moment magnitudes and canting angles.

Attempts to model the time-resolved data using the same approach used to determine the static equilibrium structure resulted in poor fits, especially for time delays shorter than 0.5 ns. This suggests that the AFM structure at such short time delays after the perturbation is more complicated than the well-ordered one found at equilibrium. More complex models of the transient magnetic structure required more parameters, which could not be constrained by the sparsity of the data measured as a function of $\Psi$.

Instead, we evaluated the nature of the transient AFM structure by comparing trends in the data to simulations of the scattered X-ray intensity. For simplicity, only changes in the relative nickel-site moments ($m_{LB} / m_{SB}$) or long-bond moment canting angles ($\theta_{LB}$) are discussed, as they were found to account for the trends observed in the measured data. Azimuthal scans of $I_\pi / I_\sigma$ were calculated for such perturbations of the static AFM structure, again following the formulation described in Appendix B. The resulting trends for different $m_{LB} / m_{SB}$ and $\theta_{LB}$ are shown in Figure 4a and 4b, respectively. It is seen that larger $m_{LB} / m_{SB}$ increases the amplitude of the two peaks in the azimuthal curve, while aligning the LB and SB moments by reducing $\theta_{LB}$ moves the peaks closer together without changing their amplitude. Figure 4c-4f also show the dependencies of the maximum and minimum $I_\pi / I_\sigma$ values obtained in an azimuthal scan. Increasing $m_{LB} / m_{SB}$ predominantly leads to an increased maximum value at +/- 90° and a slightly decreased minimum value at +/- 180°, while smaller $\theta_{LB}$ values decreases the minimum but does not affect the maximum $I_\pi / I_\sigma$ value. These different trends found for the min and max $I_\pi / I_\sigma$ suggests that changes of $m_{LB} / m_{SB}$ and $\theta_{LB}$ are in a sense orthogonal.

The maximum and minimum of $I_\pi / I_\sigma$ for each time delay extracted from the full set of measured sample orientations is shown in Figure 5a. It can already be seen from these trends that at time zero both values sharply decrease and that the maximum value recovers faster than the minimum value. Based on the simulations discussed above, these trends suggest that at early times both $m_{LB} / m_{SB}$ and $\theta_{LB}$ decrease but recover at different rates, with the relative moment following the trend in the max $I_\pi / I_\sigma$. The measured trends of the min and max $I_\pi / I_\sigma$ were then fit using the simulation trends to extract an estimation of the degree to which these parameters change throughout the structural evolution. The results of the fits are shown in Figures 5b and 5c with the error bars given by the uncertainty in the best fit parameters. The equilibrium structure shown in Figure 2 was found to lie within the confidence intervals of the recovered negative time delay parameters shown in Figure 5.

The extracted evolution of $m_{LB} / m_{SB}$ and $\theta_{LB}$, plotted in Figure 5, shows that the optical excitation affects both parameters. Most notably, the size of $m_{LB} / m_{SB}$ is found to decrease at early time delays to a value below 1, followed by a gradual nanosecond recovery back to equilibrium. This suggests that the light excitation creates a state where the LB moment becomes smaller than the SB moment, which is inverse to the spin-disproportionation found in the equilibrium low energy spin model [18,20]. This state also differs from the equilibrium metallic state, as in this case no bond disproportionation between the two nickel sites is expected.

Furthermore, Figure 5c shows that the LB spin canting angle of the light-induced magnetic state is different from the equilibrium low energy spin state. At early time delays it is found that this angle is smaller, suggesting that the LB moment is more aligned with the SB moment. However, precise quantification of this angle is difficult given the sparsity of the measurements as a function of Ψ and the corresponding uncertainties of the values refined for the canting angle. Further insight into the light induced spin order and recovery dynamics requires measurements including more azimuthal angles, which could also enable a more complex structural parameter modelling.

The observed light-induced magnetic order dynamics with inverse site spin-disproportionation could be explained by a mechanism where electronic states near the top of the valence band, associated with LB sites [32], are preferentially excited into the conduction band. This would result in a different electronic configuration on the LB sites and would disrupt the spin order by perturbing the balance of super and double exchange interactions in the material. Such a lower occupancy and higher disorder of LB-site spins would manifest as a smaller LB-site moment in the refined magnetic structure refined from resonant magnetic diffraction, which is consistent with our measurements.

In summary, we have shown that the azimuthal angle dependence of time-resolved, linearly polarized resonant X-ray diffraction measurements can be used to determine the spin ordering dynamics coupled to light-induced phase transformations. The excitation of a $NdNiO_3$ thin film by 0.61 eV femtosecond laser pulses was found to result in an abrupt reduction in the scattered intensity of a magnetic superlattice reflection. At early times in the recovery process, the magnetic structure in the film was found to differ from the equilibrium low energy spin order, having LB-sites with more collinear canting and smaller moments than SB-sites. The equilibrium AFM structure then gradually recovered on a nanosecond time scale. The observation of smaller LB- than SB-moments shortly after excitation with a photon energy near the band gap of $NdNiO_3$ suggests in the light induced state is created by preferentially exciting valence band LB states near the Fermi energy. From this result, one expects that excitation with a higher photon energy will create a metallic state without such disproportionation, as it would also involve excitations from more SB states deeper in the valence band. Thus, the nature of the light-induced metallic state may be controllable by tuning the excitation photon energy.

ACKNOWLEDGEMENTS

This work was supported by the European Research Council (ERC) grant 319286 (Q-MAC) and the Swiss National Science Foundation through Division II. Additional funding was provided by Cluster of Excellence 'CUI: Advanced Imaging of Matter' of the Deutsche Forschungsgemeinschaft (DFG) – EXC 2056 – project ID 390715994. A.S.D. was supported by a fellowship from the Alexander von Humboldt Foundation. We thank Diamond Light Source for provision of beamtime under proposal number SI17605-2.

## APPENDIX A: FILM CHARACTERIZATION

The atomic structure and electronic properties of the NdNiO$_3$ film was characterized using a set of X-ray diffraction and transport measurements.

First, the film thickness and out of plane lattice constant was determined by an X-ray scattering measurement around the 111 reflection, which was taken using a laboratory Cu K$_\alpha$ X-ray diffractometer. As shown in Figure A1a, these data agree well with a model assuming a lattice plane d-spacing of 2.202 Å and a film thickness of 36 nm [37]. Next, in-plane strain of the film was characterized by mapping the scattered X-ray intensity around the 201 reflection (Fig. A1b). As the peak intensity from the film and substrate are found at the same value for the reciprocal space in-plane direction ($Q_{ip}$), we conclude that the film was coherently strained to the LaAlO$_3$ substrate.

In addition, the metal-insulator transition in the NdNiO$_3$ film was characterized from van der Pauw resistivity measurements made on the film as a function of temperature. Figure A1c shows that below a temperature of 50 K, the resistivity of the film was found to increase by almost four orders of magnitude compared to room temperature. Such a large increase through the metal-insulator transition suggests a uniform film with correct stoichiometry [38].

## APPENDIX B: MAGNETIC SCATTERING

The resonant X-ray scattering factor from a magnetic atom is given by [39]

$$f_n = \left[(\hat{\varepsilon}' \cdot \hat{\varepsilon})F^{(0)} - i(\hat{\varepsilon}' \times \hat{\varepsilon}) \cdot \widehat{m_n} F^{(1)} + (\hat{\varepsilon}' \cdot \widehat{m_n})(\hat{\varepsilon} \cdot \widehat{m_n})F^{(2)}\right] \quad (A1)$$

where $\hat{\varepsilon}$ and $\hat{\varepsilon}'$ are the polarizations of the incident and scattered light, $\widehat{m_n}$ is the orientation vector of the magnetic moment of the n$^{th}$ ion, and $F^{(j)}$ are respective strengths of the charge, first- and second-order magnetic scattering processes. As our measurements were made on the ( ¼ ¼ ¼ ) peak, which has been shown to have no charge scattering contribution, we can ignore the first term in expression A1. Furthermore, in a specular reflection geometry, equation A1 becomes equation 15 of Ref. 36. In our case, the equilibrium antiferromagnetic cell contains four scattering planes with different moments that are uniform in the plane, as depicted in Figure 2(b). The first and third planes correspond to Ni SB-sites, while the second and fourth planes correspond to Ni LB-sites. An anti-ferromagnetic symmetry was then imposed on each pair of planes, so the second-order magnetic scattering terms in Equation A1 also cancel out. As a consequence, only the first order magnetic scattering terms from equation 15 of Ref. 36 were considered to calculate the scattered intensity.

Then, the structure factor for each scattering channel can be written as

$$F_{\sigma\sigma} = 0 \quad (A2a)$$

$$F_{\sigma\pi} = -iF^{(1)} \sum_n \left[(m_{n,x}\cos\theta + m_{n,z}\sin\theta)e^{2\pi i z_n}\right] \quad \text{(A2b)}$$

$$F_{\pi\sigma} = -iF^{(1)} \sum_n \left[(-m_{n,x}\cos\theta + m_{n,z}\sin\theta)e^{2\pi i z_n}\right] \quad \text{(A2c)}$$

$$F_{\pi\pi} = -iF^{(1)} \sum_n \left[(m_{n,y}\sin 2\theta)e^{2\pi i z_n}\right], \quad \text{(A2d)}$$

where the sum is taken over the moments in the magnetic unit cell and $2\theta$ is the X-ray scattering angle. Here, $z_n$ denotes the fractional position along $[111]_{pc}$ of the $n$-th scattering plane in the magnetic supercell. We assumed an equal spacing between all planes, corresponding to the set of positions, $z_n = \left[0, \frac{1}{4}, \frac{1}{2}, \frac{3}{4}\right]$.

The magnetic moment on the $n$-th scattering plane was defined as

$$\boldsymbol{m_n} = m_{n,x}\boldsymbol{i} + m_{n,y}\boldsymbol{j} + m_{n,z}\boldsymbol{k} = m_n[\cos(\phi_n + \Psi)\sin(\theta_n)\boldsymbol{i} + \sin(\phi_n + \Psi)\sin(\theta_n)\boldsymbol{j} + \cos(\theta_n)\boldsymbol{k}], \quad \text{(A3)}$$

where $m_n$ denotes the moment magnitude, $\phi_n$ and $\theta_n$ are orientation angles relative to the [1-10] and [111] crystal directions, and $\Psi$ is the azimuthal rotation angle of the sample. As an example, an ideal non-collinear AFM structure is then described by the angles $\phi_{SB} = 0, \theta_{SB} = 0, \pi$ and $\phi_{LB} = 0, \theta_{LB} = -\frac{\pi}{2}, \frac{\pi}{2}$.

As no polarization discriminator was used on the diffracted beam, the recorded intensities were calculated from equations A2 and A3 using

$$I_\sigma = F_{\sigma\pi}^2 \text{ and } I_\pi = F_{\pi\sigma}^2 + F_{\pi\pi}^2. \quad \text{(A4)}$$

The resulting ratio of $I_\pi / I_\sigma$ as a function of $\Psi$ was then compared to the measured data to refine the $\boldsymbol{m}_{LB}$ and $\boldsymbol{m}_{SB}$ orientations and relative magnitude using a Levenberg-Marquardt least-squares regression algorithm. The confidence interval on each parameter was calculated from the covariance matrix.

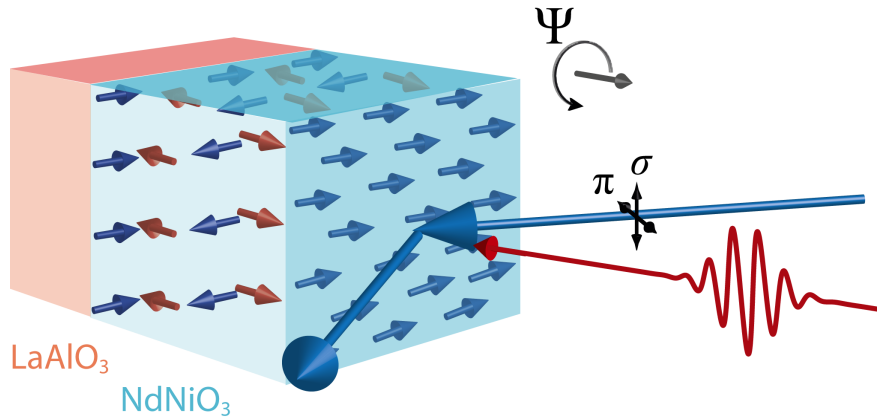

**Figure 1.** Experimental schematic showing soft X-ray (blue) and near-infrared (red) beams incident on a NdNiO$_3$ film supported by a LaAlO$_3$ substrate. The noncollinear AFM structure is oriented within the [111]$_{pc}$ NdNiO$_3$ film with the moments of the nickel long-bond (blue) and short-bond (red) sites respectively oriented in-plane and normal to the film surface. The sample azimuthal angle $\Psi$ about a vector parallel to the film surface normal, and the soft X-ray $\pi$- and $\sigma$-polarization directions are also depicted. This illustration shows the NdNiO$_3$ magnetic structure relative to the scattering plane at $\Psi = 90°$

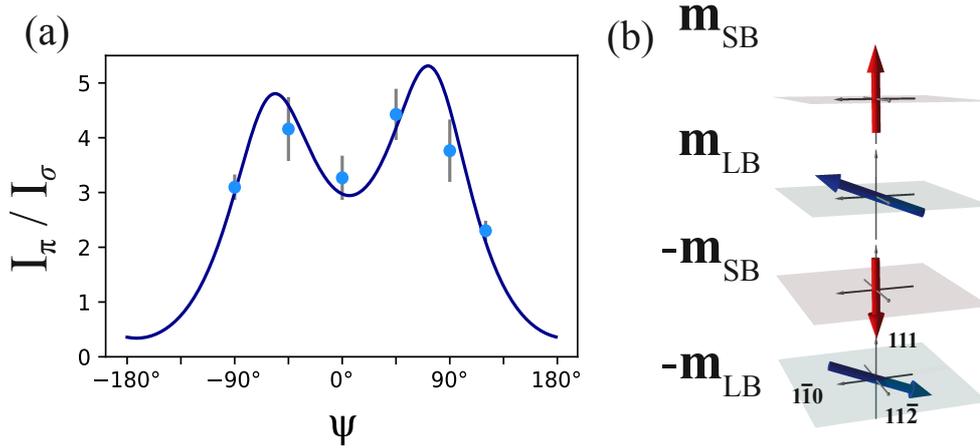

**Figure 2.** (a) Ψ-angle dependence of the (¼ ¼ ¼) AFM Bragg spot intensity at equilibrium shown in terms of the ratio of the laser-off scattered X-ray intensities measured using incident π- and σ-polarizations ($I_\pi / I_\sigma$). The data points denote the values obtained from the set of measured sample orientations, while the blue line is the trend from the AFM structure found to best fit these data. (b) The corresponding best-fit noncollinear AFM structure is illustrated, having nickel SB- and LB-moments of $0.7 \pm 0.2$ μB and $1.3 \pm 0.2$ μB respectively and a $75° \pm 5°$ canting of the LB-moment relative to the $[111]_{pc}$ direction.

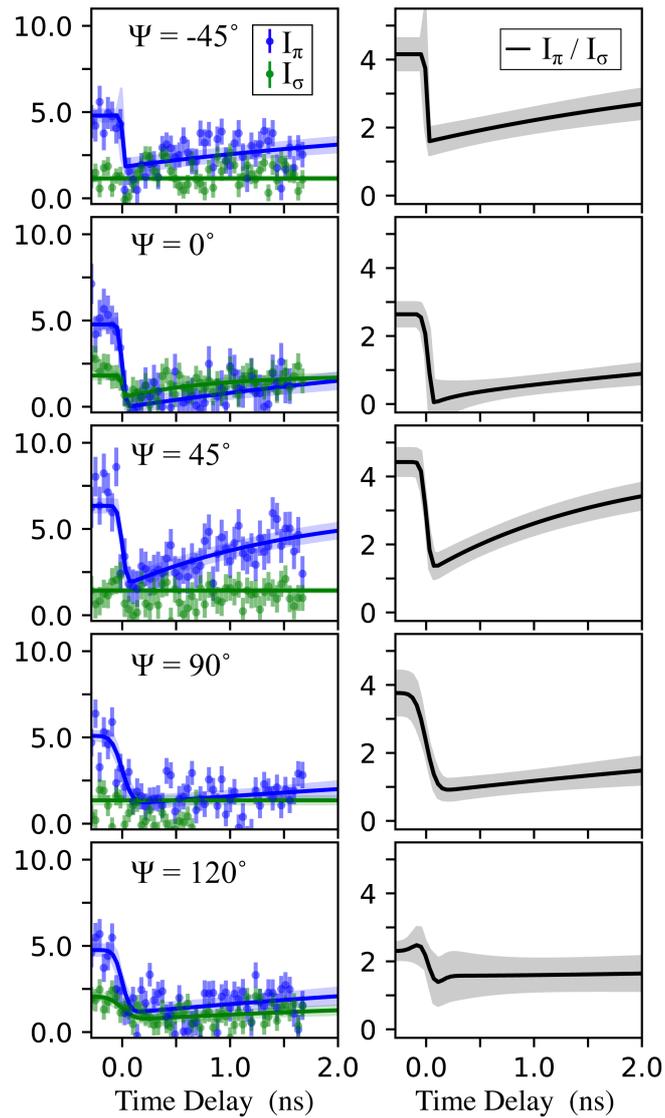

**Figure 3.** The intensity of the ¼ ¼ ¼ AFM Bragg spot as a function of time delay after the 2-μm excitation pulse for the set of measured sample azimuthal angles, Ψ, and incident X-ray polarizations is shown in the left panels. Fits to the data using an error-function with exponential decay are overlaid, including 95% confidence intervals (blue/green shaded region). The time resolved trends for the intensity ratios $I_\pi / I_\sigma$ and the confidence intervals (grey) derived from the fits are shown in the panels on the right.

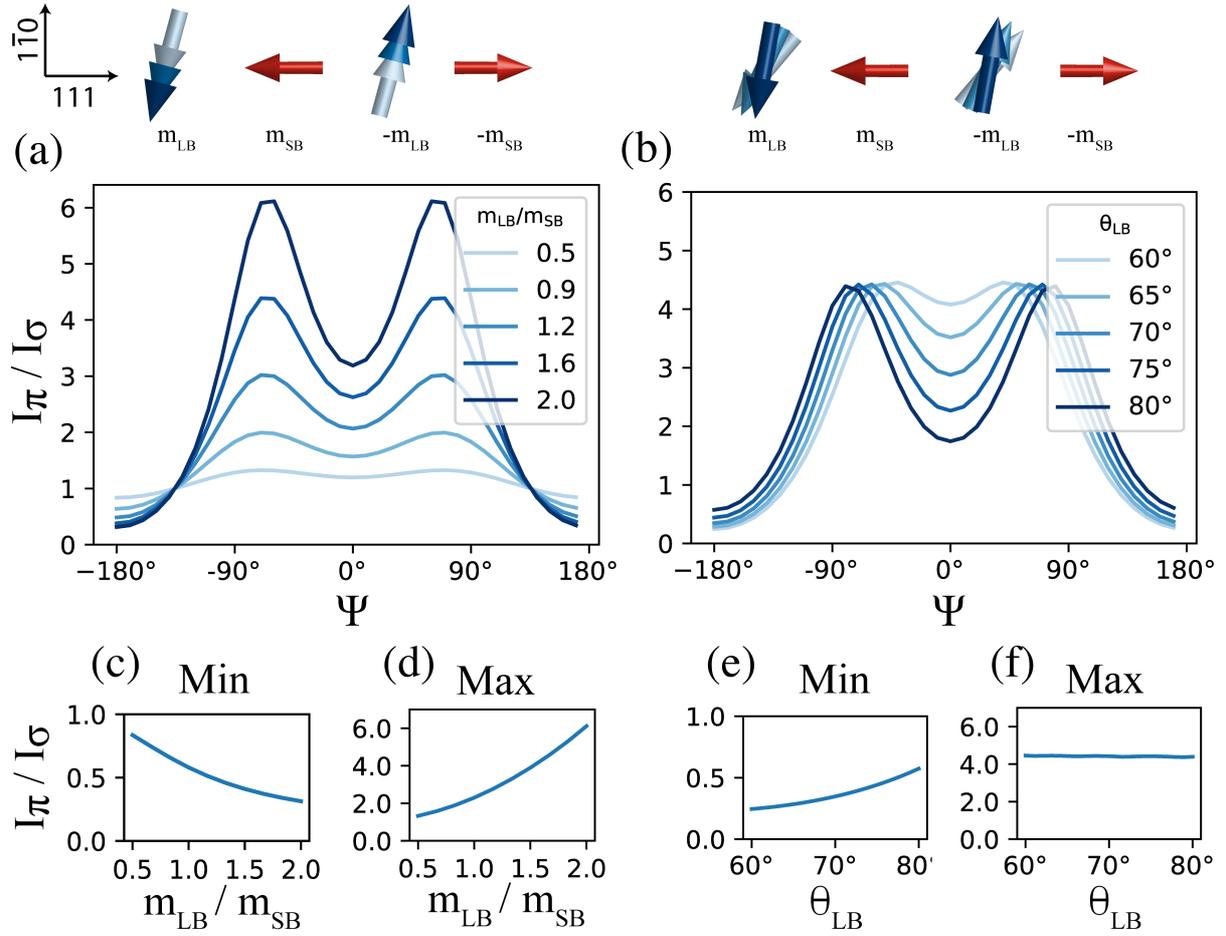

**Figure 4.** Calculated $\Psi$-dependence of the scattered intensity ratio $I_\pi / I_\sigma$, assuming magnetic structures of the NNO film with (a) different relative nickel-site moments ($m_{LB} / m_{SB}$) and (b) different long-bond site canting angles ($\theta_{LB}$). The corresponding perturbed magnetic structures are illustrated above the trends. The minimum and maximum values of $I_\pi / I_\sigma$ over the full range of $\Psi$ are shown as a function of $m_{LB} / m_{SB}$ in (c) and (d), and as a function of $\theta_{LB}$ in (e) and (f).

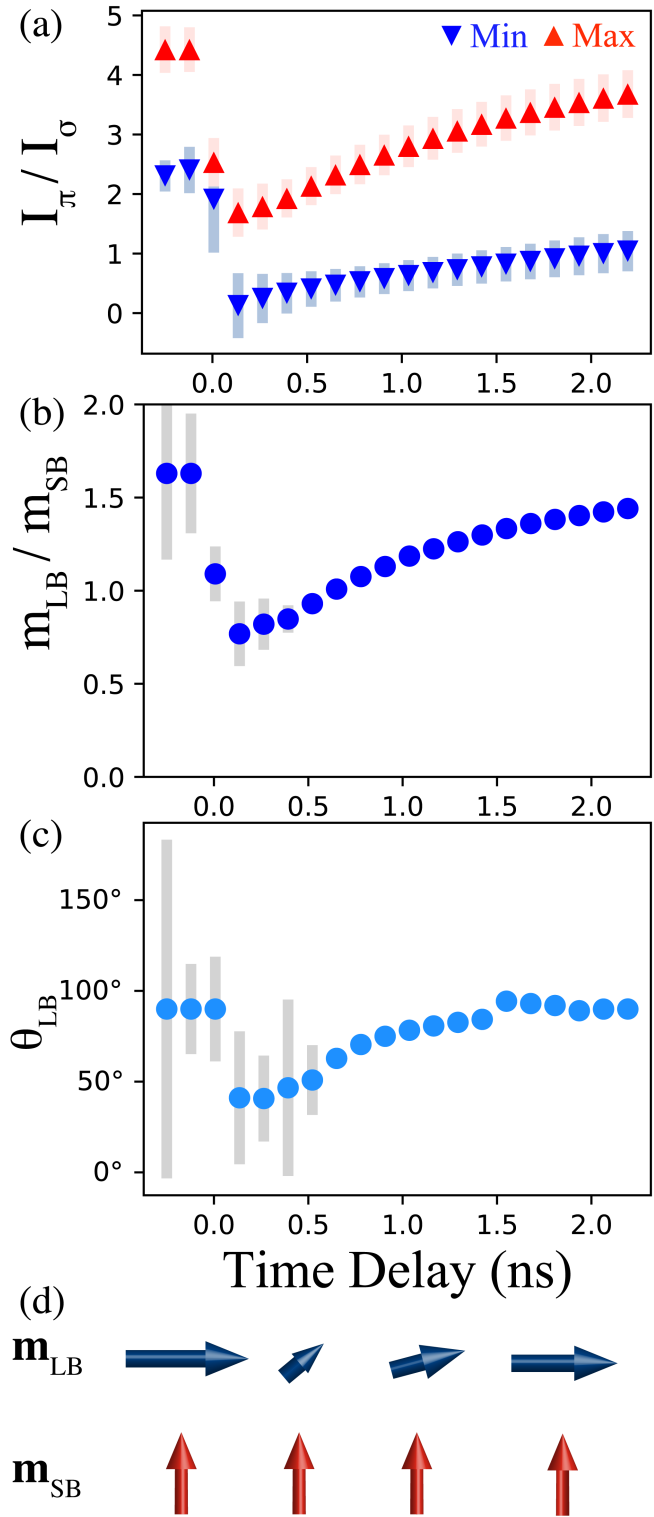

**Figure 5.** (a) Maximum and minimum intensity ratio values extracted from the set of measured sample orientations as a function of time delay. (b,c) Recovered values of the AFM structure parameters $m_{LB}/m_{SB}$ and $\theta_{LB}$ found to best fit the measured maximum and minimum. (d) The corresponding evolution of the spin order on the NdNiO$_3$ nickel sites is illustrated in register with the time axis above.

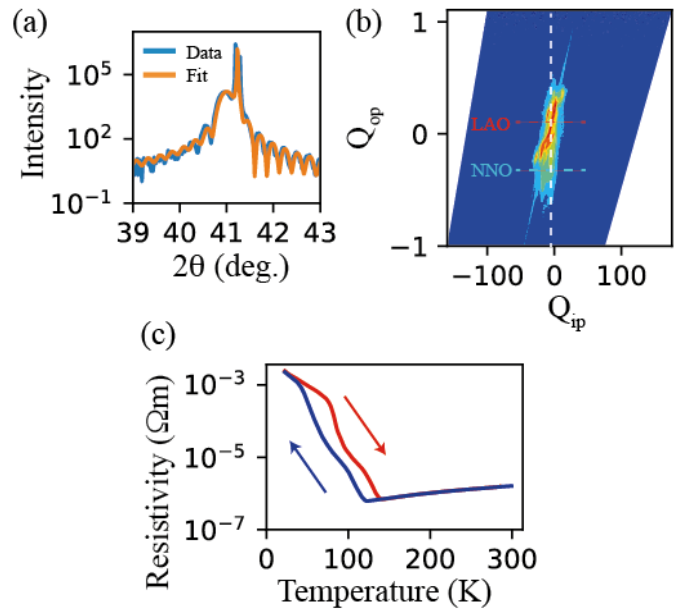

**Figure A1**. (a) A θ-2θ X-ray scattering measurement around the NdNiO$_3$ film 111 Bragg peak is overlaid by a fit assuming a film thickness of 36 nm and interplanar spacing of 2.202 Å. (b) Scattered X-ray intensity around the 201 NdNiO$_3$ Bragg reflection shown as a function of reciprocal space directions $Q_{ip}$ and $Q_{op}$, which are parallel to the film in-plane and out-of-plane directions, respectively. The blue and red dashed horizontal lines mark the scattered intensity coming from the LaAlO$_3$ substrate and NdNiO$_3$ film, while the white line is a guide to the eye at constant $Q_{ip}$. (c) The resistivity in the NdNiO$_3$ film through a temperature cycle is shown. The trend measured as the sample was cooled is shown in blue, while that while heating is shown in red.